# Evidence for exchange Dirac gap in magneto-transport of topological insulator-magnetic insulator heterostructures

S. R. Yang[1#], Y. T. Fanchiang[2#], C. C. Chen[1], C. C. Tseng[1], Y. C. Liu[1], M. X. Guo[1], M. Hong[2], S. F. Lee[3*], and J. Kwo[1*]

[1]*Department of Physics, National Tsing Hua University, Hsinchu 30013, Taiwan*

[2]*Department of Physics, National Taiwan University, Taipei 10617, Taiwan*

[3]*Institute of Physics, Academia Sinica, Taipei 11529, Taiwan*

[#] Authors who have equal contributions to this work

[*] Corresponding authors

**Abstract:**

Transport signatures of exchange gap opening because of magnetic proximity effect (MPE) are reported for bilayer structures of $Bi_2Se_3$ thin films on yttrium iron garnet (YIG) and thulium iron garnet (TmIG) of perpendicular magnetic anisotropy (PMA). Pronounced negative magnetoresistance (MR) was detected, and attributed to an emergent weak localization (WL) effect superimposing on a weak antilocalization (WAL). Thickness-dependent study shows that the WL originates from the time-reversal-symmetry breaking of topological surface states by interfacial exchange coupling. The weight of WL declined when the interfacial magnetization was aligned toward the in-plane direction, which is understood as the effect of tuning the exchange gap size by varying the perpendicular magnetization component. Importantly, magnetotransport study revealed anomalous Hall effect (AHE) of square loops and anisotropic magnetoresistance (AMR) characteristic, typifying a ferromagnetic conductor in $Bi_2Se_3$/TmIG, and the presence of an interfacial ferromagnetism driven by MPE. Coexistence of MPE-induced ferromagnetism and the finite exchange gap provides an opportunity of realizing zero magnetic-field dissipation-less transport in topological insulator/ferromagnetic insulator heterostructures.

Breaking time-reversal symmetry (TRS) in topological insulators (TIs) leads to several exotic phenomenon such as quantum anomalous Hall effect (QAHE), topological magnetoelectric effect, and magnetic monopole [1,2]. A prerequisite of these novel quantum state is an energy gap opened at the Dirac surface state induced by exchange interaction with magnetic elements [3]. Magnetic doping is a prevalent way of introducing ferromagnetism in TIs [4-7]. Study of TRS breaking in magnetically doped TIs was ignited by the direct observation of an exchange gap opening of topological surface states (TSSs) via angle-resolved photoemission spectroscopy (ARPES) [5], and culminated with the realization of QAHE in Cr-doped $(Bi,Sb)_2Te_3$ [8]. Although magnetic doping is proven to be effective in breaking TRS, the observation temperature of QAHE reported so far was less than 2 K [8-12], order-of-magnitude lower than the ferromagnetic Curie temperature ($T_C$). It is suggested that the disorder created by dopants, as well as the small exchange gap size induced by low doping concentration, poses a limit of raising the QAHE temperature [12,13].

Recently, magnetic proximity effect (MPE) of TI/ferromagnetic insulator (FI) heterostructures was demonstrated as another promising route of breaking TRS [14-17]. Besides the benefit of much higher $T_C$, the induced interfacial magnetization is uniform, free of crystal defects. A room-temperature ferromagnetism by MPE is directly observed in epitaxial $EuS/Bi_2Se_3$ by polarized neutron reflectometry [16]. Moreover, robust anomalous Hall (AH) resistances up to 400 K has been detected in $(Bi,Sb)_2Te_3$ films on TmIG with perpendicular magnetic anisotropy (PMA) [17]. Despite the clear observations of ferromagnetism and presumably pronounced TRS breaking, the experimental indications of exchange gap opening following MPE in these cases are still vague. Unlike magnetically doped-TIs where the gapped surface can be exposed to the probe of ARPES technique [5], the gapped surface state caused by MPE is buried at the interface, making it difficult to investigate using typical ARPES. Attempts to detect MPE-induced exchange gap by transport measurements have been made by various groups [18-20]. One signature of exchange gap opening is an emerging weak localization (WL) taking the form of negative magnetoresistance (MR) accompanied by a suppressed weak antilocalization (WAL) [21]. However, negative MR in TI/FI was hitherto observed in samples comparable or beyond the Ioffe-Regel limit (sheet resistance $R_s \geq h/e^2$) [18-20]. Above the limit, the Anderson (strong) localization due to disorder could also give rise to a negative MR [22,23], which cannot be easily distinguished from the one due to an exchange gap. Therefore, a definite transport signature of MPE-induced exchange gap remains elusive.

Despite the observation of negative MR and suppressed WAL [24-26], robust MPE-induced ferromagnetism in TIs, which should be best manifested as large remnant magnetization $M_r$ pointing out-of-plane and hysteric AHE, has not been observed in those systems. Conversely, although clear $M_r$ was detected in (Bi,Sb)$_2$Te$_3$/TmIG, the negative MR and transport signature of gap opening were not simultaneously presented [17]. To achieve QAHE in TI/FI, both ingredients, robust ferromagnetism of TSS and exchange gap opening, must be fulfilled concurrently, which is a condition yet to be demonstrated for TI/FI. In pursuing high-temperature QAHE through MPE, it is important to identify the transport signature of the exchange gap associated with MPE-induced ferromagnetism, before one moves on to the ultimate goal of QAHE. Moreover, although AHE is often taken as an evidence of interfacial ferromagnetism, the possible spin current effect in TI/FI, such as spin Hall MR [27], can also lead to similar AH resistances. Additional transport study is demanded to elucidate possible modulations of magneto-transport by spin current effects.

In this work we report a pronounced WL that competes with the WAL in

Bi$_2$Se$_3$/YIG and Bi$_2$Se$_3$/TmIG bilayers. The ferrimagnetic garnet films were chosen because of their high $T_C$ above 500 K [28], good thermal stability in conjunction with TIs, and their technological importance [29]. The emergent WL effect is strong enough to manifest a negative magnetoresistance (MR), showing systematic changes with the perpendicular magnetization controlled by tilted external fields. This strongly suggests that the observed WL arose from a finite MPE-induced exchange gap, whose size could be further tuned by the external field directions. Most importantly, the breaking of TRS by MPE is corroborated by the observation of AHE up to 180 K, together with clear anisotropic magnetoresistance (AMR), confirming ferromagnetic TSS. Our study thus presents a coherent picture of the long sought MPE-induced ferromagnetism and exchange gap in electron transport.

Our YIG and TmIG thin films are fabricated using off-axis sputtering [30,31]. YIG is a ferrimagnetic insulator with ultralow magnetic damping ideal for spin wave and spin current transport study [29], while TmIG films, when under a tensile strain, exhibit robust and tunable perpendicular magnetic anisotropy (PMA) [31-33] which is essential for exchange gap of TSSs. Bi$_2$Se$_3$ thin films of 6 – 40 quintuple layer (QL) were deposited on the garnet substrates by molecular beam epitaxy (MBE). High quality Bi$_2$Se$_3$ thin films and sharp Bi$_2$Se$_3$/garnet interface were obtained by the invention of a novel growth procedure, a key factor for conveying strong exchange interaction of the localized magnetic moments of garnets layers and TSSs [34]. For transport measurements, Bi$_2$Se$_3$/garnet bilayer samples were made into Hall bars (650 μm × 50 μm) by standard photolithography. Four points measurement was carried out in a 9 T Quantum Design physical property measurement system with a 10 μA DC current.

Figure 1(a) shows the temperature dependence of sheet resistance ($R_s$) of Bi$_2$Se$_3$/YIG bilayers with Bi$_2$Se$_3$ thickness of 6, 10, 16, and 40 QL. All samples exhibit metallic behavior of decreasing $R_s$ when the samples were cooled down from room temperature. The sheet carrier density of these samples is in the range of $(1.5 - 3) \times 10^{13} \text{cm}^{-2}$, indicating that the bulk carriers of $Bi_2Se_3$ participate in the electron transport. Due to an increasing surface scattering, the $R_s$ tends to be larger in thinner $Bi_2Se_3$ [35]. Note that the maximum $R_s$ of these sample were well below $h/e^2 (\approx 25.8 \text{ k}\Omega)$ satisfying the condition of transport regime, thus quantum interference effects of 2D electron systems in TI thin films, such as WAL and possible WL, can be described by well-developed theories [22]. Figure 1(b) displays the MR data taken at 2 K under a perpendicular applied field for the four $Bi_2Se_3$/YIG bilayers and, for comparison, a 9 QL $Bi_2Se_3$ grown on $Al_2O_3$. For $Bi_2Se_3/Al_2O_3$, a sharp cusp feature at low fields, characteristic of WAL effect, was observed, and the MR stayed positive up to 9 T. WAL in thin $Bi_2Se_3$ is generally attributed to destructive interference because of difference of Berry's phase $2\pi$ accumulated by the Dirac fermion travelling in two time-reversed paths [1]. In contrast, notable negative MR were observed in 6 and 10 QL $Bi_2Se_3$/YIG. Specifically, at low fields ($< 1$ T), a weakened positive MR or suppressed WAL was observed in all the four $Bi_2Se_3$/YIG bilayers. At intermediate fields ($1 - 4$ T), MR becomes negative for the thinner two samples. While the MR of 16 and 40 QL $Bi_2Se_3$/YIG remained positive, that from 16 QL $Bi_2Se_3$/YIG did show the much weaker positive MR compared to that of $Bi_2Se_3$/sapphire. When the external field exceeded 4 T, all the samples exhibited positive MR that results from the Lorentz force on moving electrons.

The suppressed WAL effect of $Bi_2Se_3$/YIG suggests that an additional transport mechanism shows up that contributes to a negative MR and competes with the WAL. Since the film quality of $Bi_2Se_3$ grown on YIG is comparable to that on sapphire, the distinct MR behavior of $Bi_2Se_3$/YIG is most likely originated from the interaction between the bottom surface of $Bi_2Se_3$ and the YIG layer. As the YIG is ferrimagnetic at 2 K [36], a sizable interfacial exchange coupling should exist in $Bi_2Se_3$/YIG [15,25,30], which is otherwise negligible for the non-magnetic sapphire substrate. Therefore, the stark contrast in the MR behaviors suggest that the suppressed WAL and negative MR are the indication on transport properties of TRS-breaking in TIs. Other possible mechanisms for the negative MR of TIs are defect-induced hopping transport [37], hybridization gap of TSSs [21], and bulk subbands in thin TIs [38]. Each scenario can be excluded straightforwardly, as discussed in details in Supplemental Materials [39].

When the TSS is subjected to an exchange field, the Dirac fermion becomes massive, as expressed by an effective Hamiltonian $H = -i\hbar v_F(\hat{\sigma} \times \hat{z}) \cdot \nabla + \frac{J\hat{\sigma}}{2} \cdot \mathbf{M}$ [21], where $v_F$ is the Fermi velocity, $\hat{\sigma}$ is Pauli matrix, $J$ is the exchange coupling constant and $\mathbf{M}$ is the magnetization unit factor. The resulting energy dispersion is

$$E = \pm \sqrt{\left(\hbar v_F k_x + \frac{J}{2}M_y\right)^2 + \left(\hbar v_F k_y - \frac{J}{2}M_x\right)^2 + \left(\frac{J}{2}M_z\right)^2} \qquad (1)$$

with an exchange gap size of $JM_z$. It has been shown that electrons travelling a closed path would acquire a Berry's phase as $\pi(1 - JM_z/2E_F)$, where $E_F$ is Fermi energy measured from the Dirac point [21]. The modulated Berry's phase weakens the associated destructive interference, or even induces a crossover from WAL to WL. In the case of $Bi_2Se_3$/YIG, interfacial exchange coupling can induce a finite gap through MPE, which further leads to competing WL. To quantitatively describe the negative MR, we calculated the longitudinal conductivity $G_{xx}$ from the tensor relation $R_{xx}/(R_{xx}^2 + R_{yx}^2)$. The competition between WAL and WL can be described using the modified Hikami-Larkin-Nagaoka (HLN) equation [21,40],

$$\Delta G = G_{xx}(B) - G_{xx}(0) = \sum_{i=0}^{1} \alpha_i \left(\frac{e^2}{\pi h}\right) \left[\psi\left(\frac{\hbar}{4el_i^2 B} + \frac{1}{2}\right) - \ln\left(\frac{\hbar}{4el_i^2 B}\right)\right] + \beta B^2 \quad (2)$$

, where $\psi$ is the digamma function, $\alpha_i$ represents the weights of WL (i = 0) or WAL (i = 1), $l_i$ is the corresponding effective phase coherence length. The $\beta B^2$ term primarily results from the Lorentz deflection of carriers [39]. To clearly reveal the presence of the WL component, the MC curves subtracted by the $\beta B^2$ background are plotted in Fig. 1(c). Positive MC was observed for 6, 10, and 16 QL Bi$_2$Se$_3$/YIG. Figure 1(d) shows the Bi$_2$Se$_3$ thickness dependence of $\alpha_0$ and $\alpha_1$. For 6 and 10 QL Bi$_2$Se$_3$/YIG bilayers, large $\alpha_0$ values of 0.7 and 2.7 were extracted, respectively. A crossing of the magnitudes of $\alpha_0$ and $\alpha_1$ was observed in thicker Bi$_2$Se$_3$ as $\alpha_0$ decreased substantially for 16 QL Bi$_2$Se$_3$ and became vanishing for 40 QL Bi$_2$Se$_3$. Meanwhile, the $\alpha_1$ value in general remains lower than -0.5, showing a slight decrease toward thinner Bi$_2$Se$_3$. The smaller $\alpha_1$ value suggests suppressed WAL channels in the bulk-surface-coupled Bi$_2$Se$_3$ [23,35]. The extracted $\alpha_0$'s and $\alpha_1$'s thus reveal the competitive behavior between WL and WAL, whose interfacial origin are indicated by the stronger WL and weaker WAL in thinner Bi$_2$Se$_3$. It is noteworthy that Eq. 2 is derived from a model considering an effective Hamiltonian of a single Dirac surface state. In reality, the Bi$_2$Se$_3$ films have two conducting surfaces interacting through the bulk carriers in transport [23]. The complexity may give rise to the unexpectedly large $\alpha_0$ in thin Bi$_2$Se$_3$/YIG. Nevertheless, Eq. 2 and Fig. 1(d) do capture the concept of emergent WL from TRS breaking at the interfaces [13].

TmIG films with PMA are more desirable for exchange gap opening at zero applied field due to their robust $M_r$. To further verify the relation between WL and exchange gap, we tilted the applied field from the z direction. Based on Eq. (1), tilted field should effectively vary the size of $M_z$ and thus tune the exchange gap size $JM_z$.

In Bi$_2$Se$_3$/TmIG, $M_z$ stands for the z component of (i) magnetization of TmIG near the interface $\boldsymbol{m}_0$ or (ii) MPE-driven magnetization on the TI side $\boldsymbol{m}_1$. Figure 2(c)-(f) shows the MR results of Bi$_2$Se$_3$/Al$_2$O$_3$ and Bi$_2$Se$_3$/YIG under applied fields of different angles $\theta_{yz} = 90°, 60°,$ and $30°$. For Bi$_2$Se$_3$/Al$_2$O$_3$, although the MR curves change with $\theta_{yz}$ in Fig. 2(c), when plotted as a function of perpendicular field $B_z (= B\sin\theta_{yz})$ in Fig. 2(d), the curves collapse into one. The observation implies that MR here is sensitive to $B_z$ only and can be well explained by an ordinary WAL effect in the Bi$_2$Se$_3$/Al$_2$O$_3$ where $J \approx 0$.

In sharp contrast, unusual MR behaviors were seen for Bi$_2$Se$_3$/TmIG. Firstly, Bi$_2$Se$_3$/TmIG also shows clear negative MR in the intermediate fields as Bi$_2$Se$_3$/YIG (Fig. 2(e)). Note that it is difficult to directly measure the $B$-dependent magnetization of TmIG films because of large low-$T$ paramagnetic background from GGG [30,41]. The total anisotropy field of TmIG is ~0.07 T at room temperature [31], which should not increase dramatically at low $T$ as it is compromised by increasing saturation magnetization of TmIG. At fields $B > 1.2\,T$ where negative MR starts to appear, we expect that the magnetization of TmIG has been saturated by $B$. Secondly, as shown in Fig. 2(f), the $B_z$ dependence of MR systematically changes with $\theta_{yz}$. At low $B_z$, the MR curves for different $\theta_{yz}$'s coincide well because WAL is governed primarily by $B_z$. The MR curves split when $B_z > 1.2\,T$ and possess weaker negative MR for smaller $\theta_{yz}$. The correspondence between negative MR and $\theta_{yz}$ can be best explained by a tunable exchange gap. As illustrated in Fig. 2(a), when the applied field was sufficient to align **M** at interface, the exchange gap size is tuned by re-orienting **M**. Since the exchange gap $JM_z \propto M_z \propto \sin\theta_{yz}$, it follows that the negative MR of Bi$_2$Se$_3$/TmIG, or the weight of WL, is in positive correlation with the exchange gap size. Because the exchange gap size determines the deviations of the Berry's phase from $\pi$, the effect of gap tuning is manifested as the variable negative

MR with $\theta_{yz}$.

In principle, an exchange gap can be induced *locally* by individual magnetic impurities [42]. Although magnetic impurities deposited on a TI surface can acquire ferromagnetism via Ruderman–Kittel–Kasuya–Yosidas (RKKY) type interaction mediated by Dirac fermions [42], this may not be the leading mechanism for ferromagnetism in a TI/FI system, where interlayer exchange coupling plays the major role. To realize QAHE, ferromagnetism needs be established for an exchange gap opened *macroscopically* without an applied field [13]. In the following, we show that the Bi$_2$Se$_3$/TmIG does meet the criterion. Figure 3(a) shows a representative curve of AHE at 100 K. A square hysteresis loop of Hall resistance was observed after the contribution from the ordinary Hall effect was subtracted, based on the empirical formula $R_{yx} = R_H(B) + R_{AH}(B)$, where $R_H$ is the ordinary Hall resistance, and $R_{AH}$ represents the AH resistance. Since TmIG layer is insulating, the AH resistance dominantly comes from the TI layer. The hysteresis loop resembles that of TmIG magnetization [31]. As displayed in Fig. 3(b), the switching field of the hysteresis loops $B_c$ increases rapidly as temperature was lowered. The enhanced $H_c$ is likely associated with the larger strain in TmIG at low temperatures [33]. Moreover, the effect of the stray field on $R_{AH}$ was negligible as we did not observe an AH resistance in Bi$_2$Se$_3$/Al$_2$O$_3$/TmIG, in which the interfacial exchange coupling is greatly suppressed by nonmagnetic Al$_2$O$_3$ (see Supplemental materials, Fig. S2(a), (b) [39]). The above observations indicate that a spontaneous magnetization, $\boldsymbol{m}_1$, has developed at Bi$_2$Se$_3$/TmIG interface because of MPE, with the magnetized Bi$_2$Se$_3$ bottom surface effectively acting as a magnetic conductor. The AH resistance can be further transformed to AH conductance using $\sigma_{AH} \cong \rho_{AH}/\rho_{xx}$. Figure 3(b) shows the temperature dependence of AH conductance amplitude $\sigma_{AH}$. $\sigma_{AH}$ decays moderately with increasing $T$ below 50 K, and persists up to 180 K. In the weakly disorder limit, the exchange gap size can be estimated from total $\sigma_{AH}$ using $\sigma_{AH} \approx \frac{e^2}{h}\left(\frac{J}{E_F}\right)^3$, taking into account of extrinsic AH conductivity [43]. With $E_F > 0.15$ eV for bulk-conductive Bi$_2$Se$_3$ [23], a lower bound of exchange gap size $\sim 7.7$ meV at 10 K is determined. The gap size is in good agreement with 9 meV obtained from density-functional theory calculations for EuS/Bi$_2$Se$_3$ [44]. The gap is one order of magnitude smaller than that observed in magnetically doped TIs of $\sim 100$ meV [13,45]. However, the very large surface state gap in doped TIs is likely nonmagnetic and caused by resonant states induced by impurities near the Dirac point [45].

To clarify the role of MPE-induced magnetization, the MR measurements were conducted at 100 K to preclude low-field quantum interference effects of TSS. Figure 4(a) shows the field-dependent resistance of our samples with longitudinal ($R_\parallel$), transverse ($R_T$), and perpendicular ($R_\perp$) fields. Distinct turnings of $R_\parallel$ and $R_T$ were observed at $\sim 0.5$ T, which were attributed to the field needed to fully saturate the perpendicular magnetization in Bi$_2$Se$_3$/TmIG toward in-plane direction. Below the saturation field, $R_\parallel$ and $R_T$ progressively increase with the increasing in-plane field, and $R_\parallel > R_T$ in particular. Meanwhile, $R_\perp$ is parabolic because of $B$-induced Lorentz deflection (see also Figs. 4(c) and (d)). We recognize the MR behaviors in Fig. 4(a) as features of AMR caused by MPE. Regardless of the domain configuration, an in-plane field promotes (diminishes) the average of in-plane (perpendicular) MPE-induced magnetization $\langle m_{1x,1y}\rangle$ ($\langle m_{1z}\rangle$) until the saturation field was reached. The subsequent increase of $R_\parallel$ and $R_T$ implies that $\langle m_{1x}\rangle$ and $\langle m_{1y}\rangle$ contribute larger resistances than $\langle m_{1z}\rangle$ does, i.e. $R_\parallel, R_T > R_\perp$. Indeed, in the regime $|B| < 0.7\ T$ where $R_\perp$ was not overwhelmingly enhanced by $B$, the AMR relation of magnetic thin films, $R_\parallel > R_T > R_\perp$ [46], has already appeared. Furthermore, Figure 4(a) also rules out SMR to be the dominant source of the AH resistances because SMR features $R_\parallel \approx R_\perp > R_T$ [27].

The AMR amplitude $R_\parallel - R_T$ continues to build up as $B$ went larger, further justified by the $(\cos)^2$ dependence on $\phi_{xy}$ shown in Fig. 4(b). Corresponding planar Hall effect characteristic of ferromagnetic conductors was also detected (Supplemental materials, Fig. S4(a), (b) [39]). Alternatively, $R_\parallel - R_T$ can be extracted from the resistance difference of $\theta_{yz}$ and $\alpha_{xz}$ scans displayed in Fig. 4(c), despite the large contributions of Lorentz deflection in $R_\perp$. As a comparison, nearly no $\phi_{xy}$ dependence of resistances was detected in $Bi_2Se_3/Al_2O_3$ (Fig. 4(d)). We notice that the field-enhanced $R_\parallel - R_T$ was also reported in Pt/YIG, where the authors stated that the large-field $R_\parallel - R_T$ mostly came from the "hybrid MR" of MPE [47], which exhibits the same relation $R_\parallel \approx R_\perp > R_T$ as that of SMR. Notwithstanding the similarity between the two systems, we point out that $Bi_2Se_3$ cannot be simply treated as a heavy metal with strong spin-orbit-coupling even at an elevated temperature. The MPE in $Bi_2Se_3$/TmIG involves hybridization between Fe *d*-orbitals and the paramagnetic TSS arising from the Bi and Se *p*-orbitals, as opposed to Pt/YIG or other Pt/FM structures where *d-d* interaction is responsible for MPE. The distinct MR behaviors, AMR in $Bi_2Se_3$/TmIG contrary to SMR/hybrid MR in Pt/YIG, may be an important clue to the microscopic transport property of MPE in TI/FI.

To summarize, a competing WL along with a suppressed WAL has been observed in $Bi_2Se_3$/YIG and $Bi_2Se_3$/TmIG. $Bi_2Se_3$ thickness dependence study suggests that the WL comes from an exchange gap of TSSs opened at interface. In addition, the weight of WL evolves with tilted MPE-induced magnetization. Such angular dependence consolidates the exchange gap as the origin of WL, and the variable WL strength signifies tunability of the gap. Moreover, the well-defined square $R_{AH}$ loops in $Bi_2Se_3$/TmIG unambiguously point to a long-range ferromagnetic order at the interface, and thus ensure a macroscopic and uniform exchange gap at zero field. The MPE-induced ferromagnetism in $Bi_2Se_3$/TmIG is doubly evidenced by typical AMR characteristics, alleviating the concern of spin current effects on the magneto-transport. The simultaneous presence of the MPE-induced long-range ferromagnetic order at the surface and the exchange gap is thus realized in the prototypical TI $Bi_2Se_3$, pending the Fermi-level tuning to deplete the bulk conduction. Lastly, by circumventing the inhomogeneous magnetic doping and the impurity-induced resonance state problems that have been encountered in magnetically doped $Bi_2Se_3$ [13,45], our study demonstrates that MPE could be a more viable way of introducing ferromagnetism in various TI systems.

**Acknowledgment**

Technical support from Nano group Public Laboratory, Institute of Physics, at Academia Sinica in Taiwan, is acknowledged. The work is supported by MoST 105-2112-M-007-014-MY3, 106-2112-M-002-010-, 107-2622-8-002-018, and 105-2112-M-001-031-MY3 of the Ministry of Science and Technology in Taiwan.

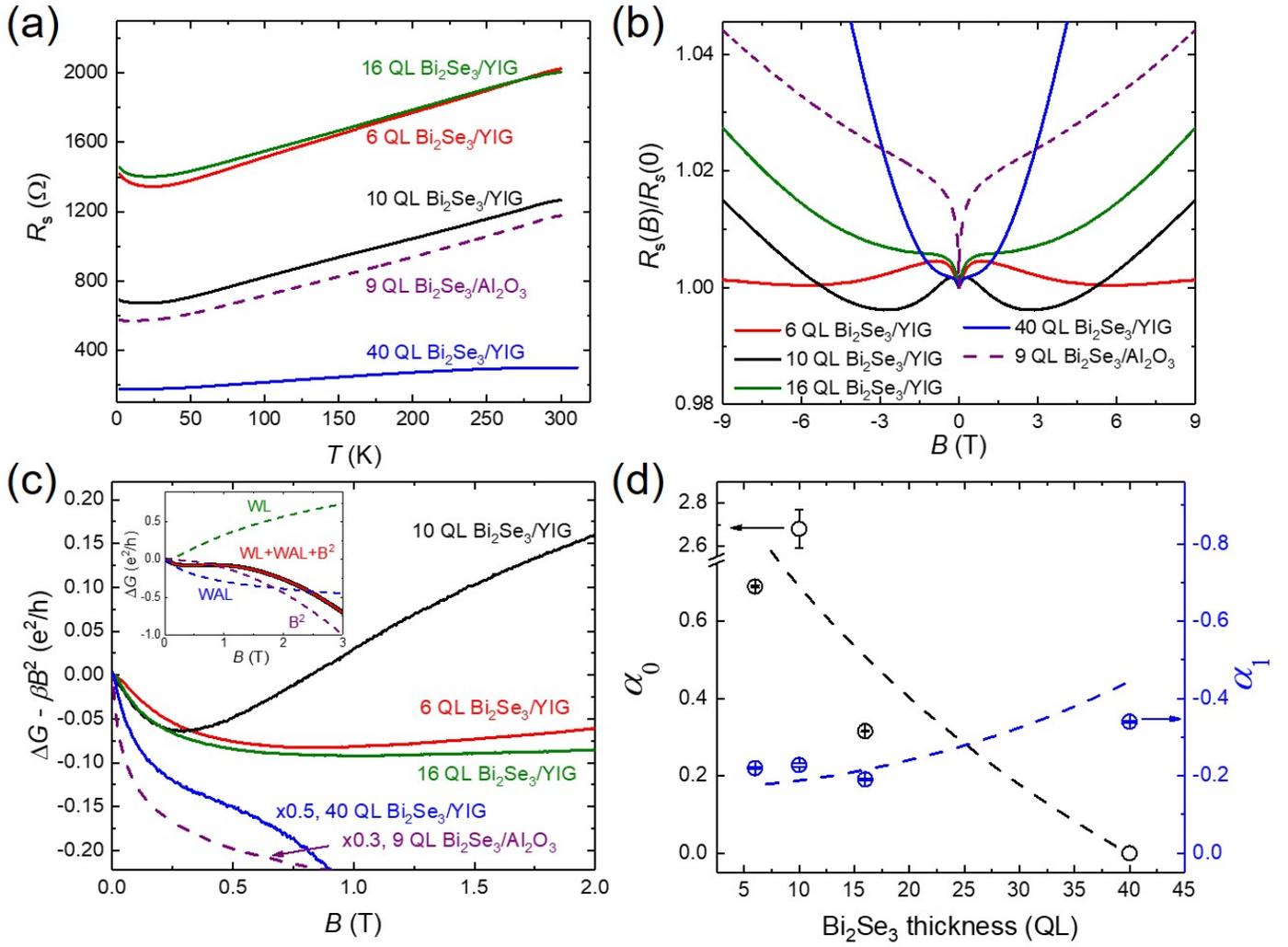

Figure 1. Transport properties of Bi$_2$Se$_3$/YIG of various Bi$_2$Se$_3$ thickness and one Bi$_2$Se$_3$/Al$_2$O$_3$ bilayer. (a) Sheet resistance $R_s$ vs temperature $T$. (b) Magnetoresistance (MR) measured at 2 K. (c) The magnetoconductance (MC) obtained by subtracting the contribution from the $\beta B^2$ term of Eq. (2). Inset: decomposition of the MC curve into the WL, the WAL, and the B$^2$ components for the 10 nm Bi$_2$Se$_3$/YIG samples. The MC curve can be well-fitted to Eq. (2). (d) Thickness dependence of $\alpha_0$ and $\alpha_1$ extracted by curve fitting to Eq. (2).

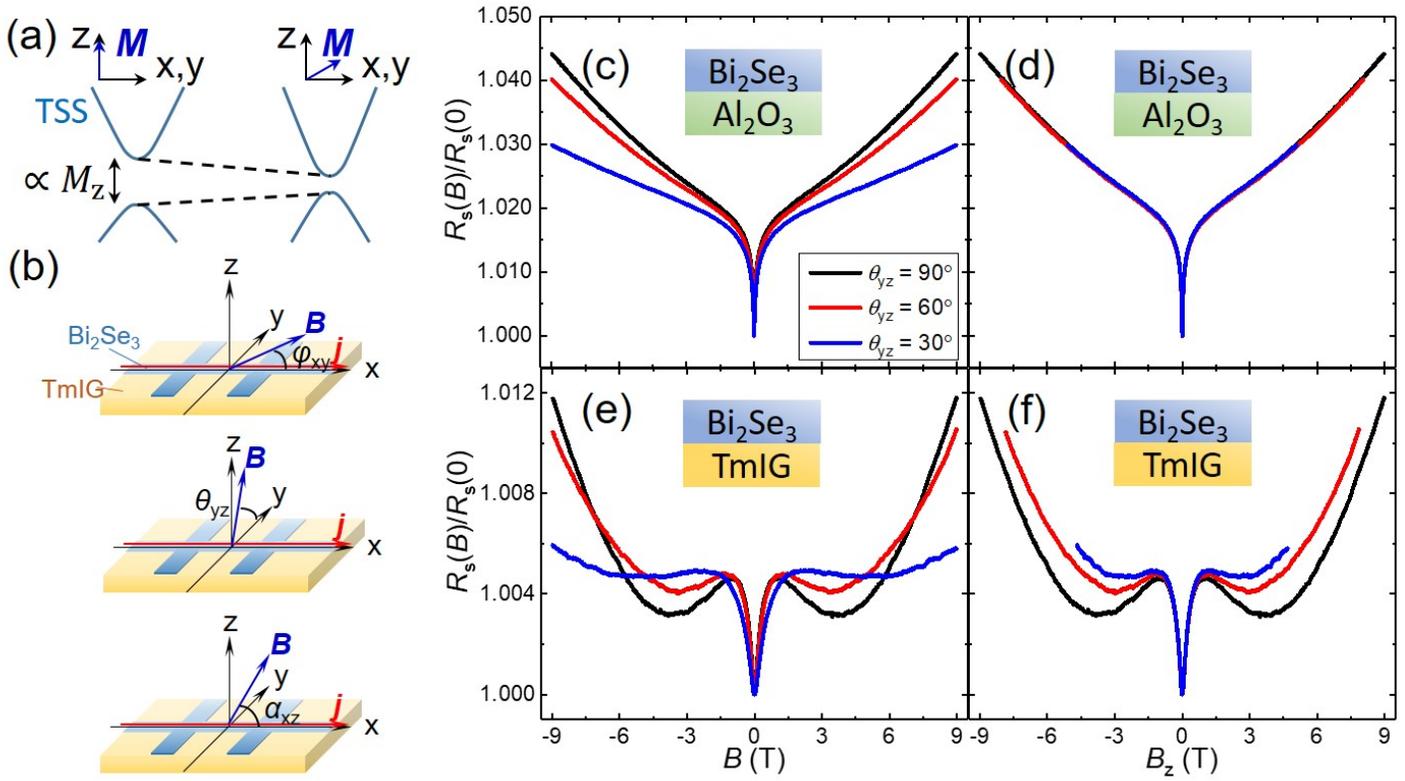

Figure 2. (a) Illustration of exchange gap opening and its size dependence on the direction of $M$. (b) Configurations of three different angular dependent resistance with field rotating in the xy-plane ($\phi_{xy}$-scan), yz-plane ($\theta_{yz}$-scan), and xz-plane ($\alpha_{xz}$-scan). (e)-(f) MR measurement at 2 K with magnetic field applied at $\theta_{yz} = 90°, 60°$ and $30°$. In (c) and (e), resistances are plotted as a function of the magnetic field strength for $Bi_2Se_3/Al_2O_3$ and $Bi_2Se_3/TmIG$, respectively. The field data are further transformed by $B\sin\theta_{yz} = B_z$ to show the $B_z$ dependence of MR in (d) and (f).

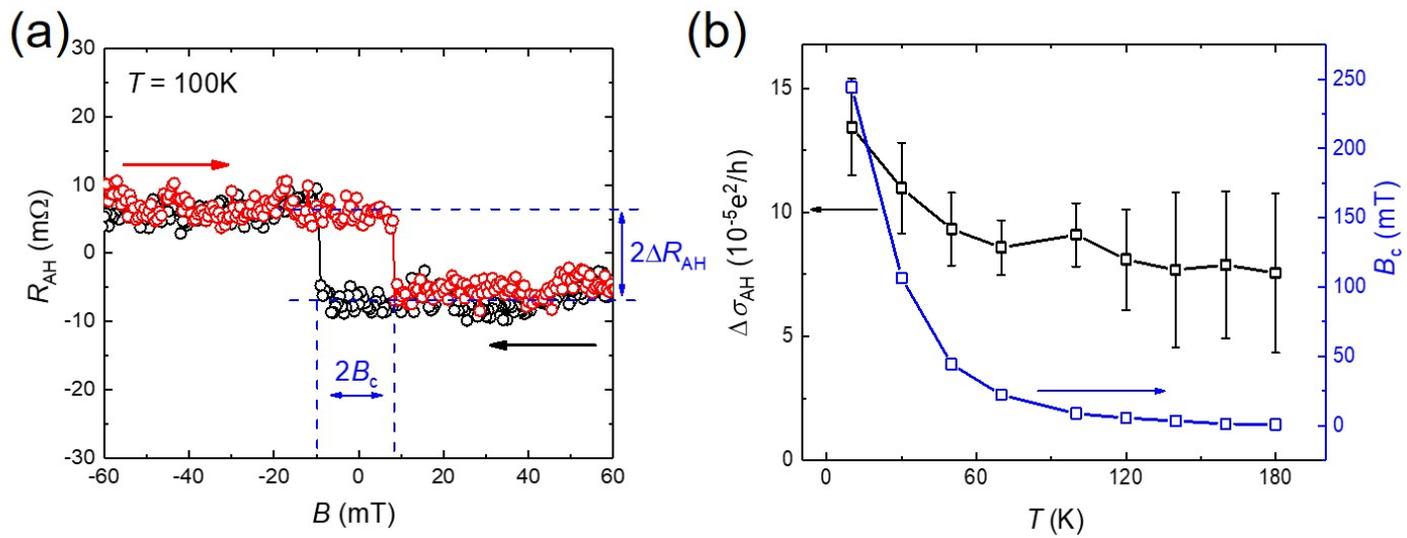

Figure 3. (a) A representative $R_{AH} - B$ hysteresis loop of $Bi_2Se_3$/TmIG at 100 K. (b) Amplitude of AH conductance $\Delta\sigma_{AH}$ and $B_c$ as a function of temperature.

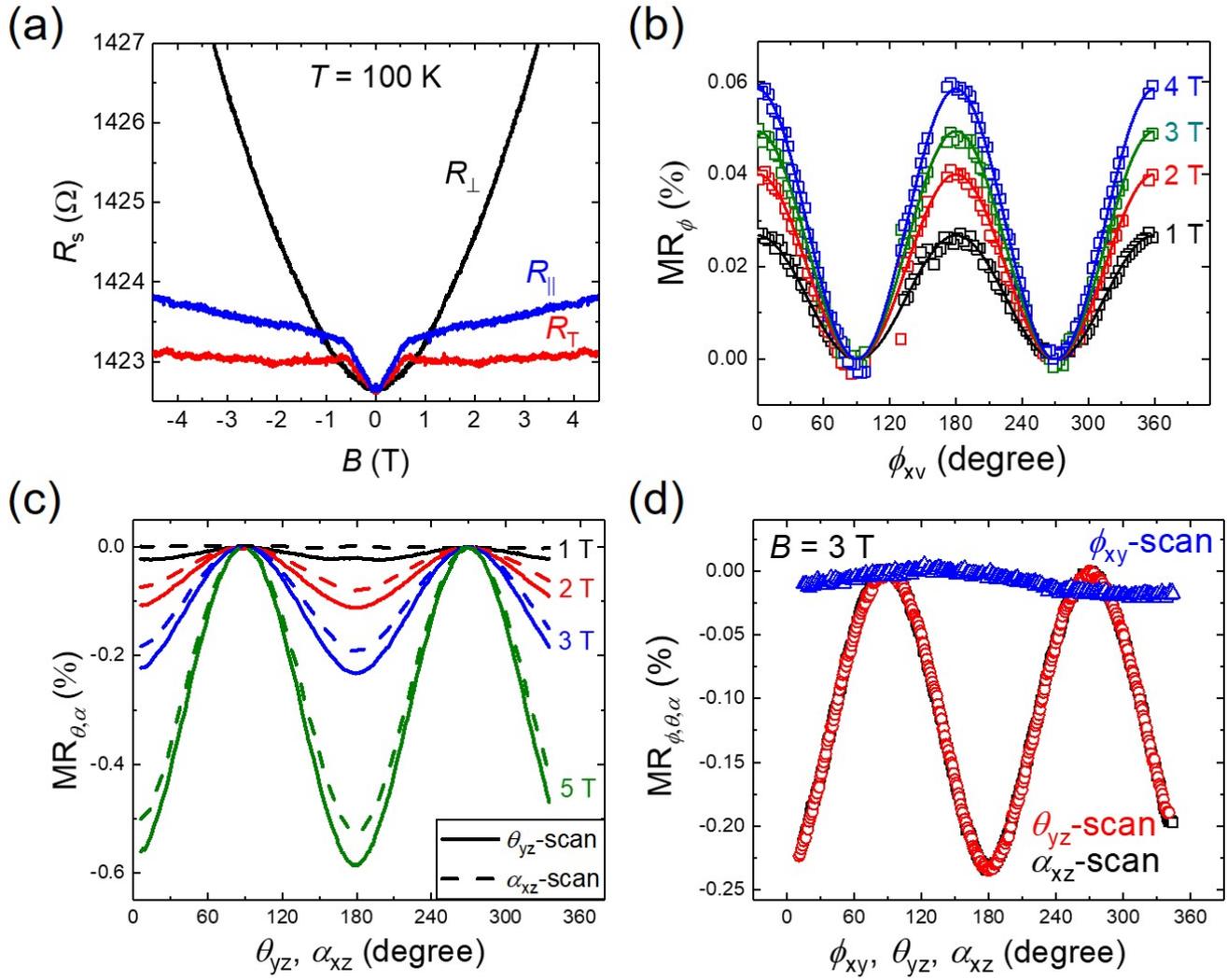

Figure 4. Field- and angular dependent resistances of $Bi_2Se_3$/TmIG and $Bi_2Se_3$/$Al_2O_3$ at 100 K. (a) Field-dependent $R_\parallel$, $R_T$, and $R_\perp$ of $Bi_2Se_3$/TmIG. (b) $\phi_{xy}$-, (c) $\theta_{yz}$-, and $\alpha_{xz}$-dependent resistances of $Bi_2Se_3$/TmIG. Here $MR_i$ is defined as $(R_s(i) - R_s(90°))/R_s(90°)$ with $i = \phi_{xy}, \theta_{yz}$, and $\alpha_{xz}$. (d) $\phi_{xy}$-, $\theta_{yz}$-, and $\alpha_{xz}$-dependent resistances of $Bi_2Se_3$/$Al_2O_3$.

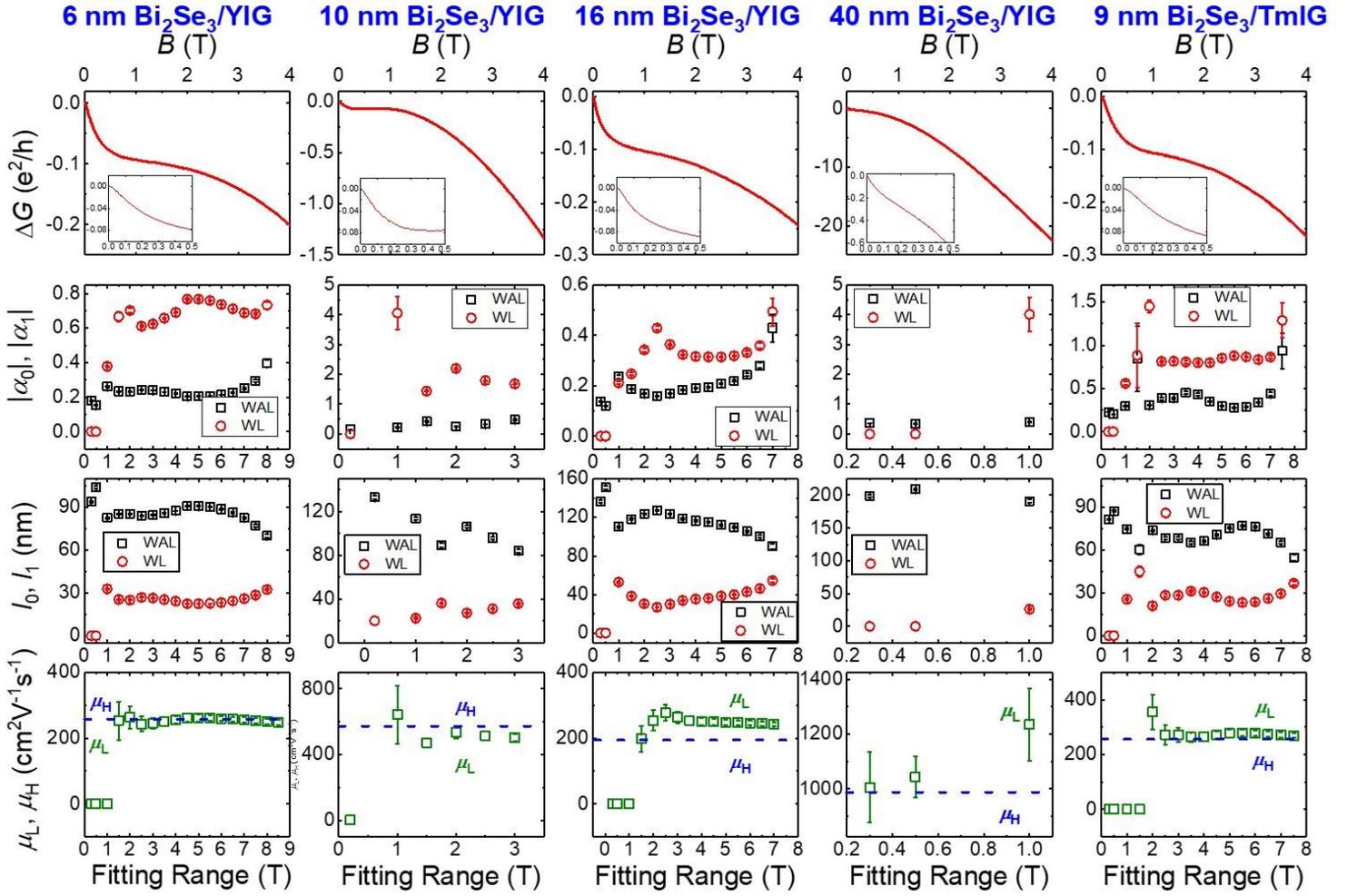

Fig. S1. Dependence of the fitted parameters on the curve fitting range. The figures in each column display data belonging to a specific sample indicated by the column headlines. The top row displays the MC curves of each sample, where the insets show the MC curves at $B < 0.5$ T. The 2nd, 3rd, and 4th rows show the extracted $\alpha_0$ and $\alpha_1$, $l_0$ and $l_1$, and the electron mobility obtained by curve fitting ($\mu_L$) and Hall measurements ($\mu_H$), respectively. Error bars represent the standard errors of the fitted parameters.

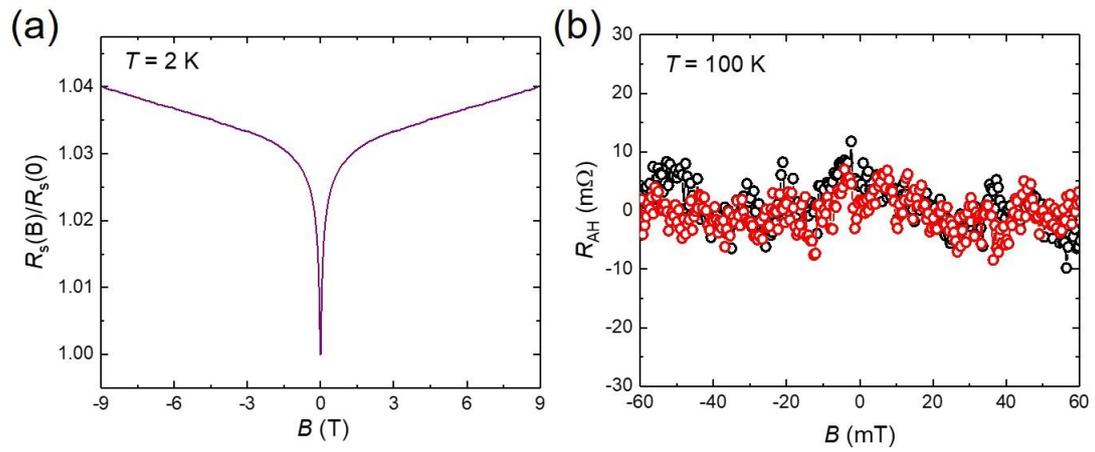

Fig. S2. Magneto-transport data of 6 nm $Bi_2Se_3$/3 nm $Al_2O_3$/TmIG. (a) The MR curve measured at 2 K. (b) The anomalous Hall resistance $R_{AH}$ taken at 100 K. Black and redy symbols represent magnetic field sweeping up and down, respectively.

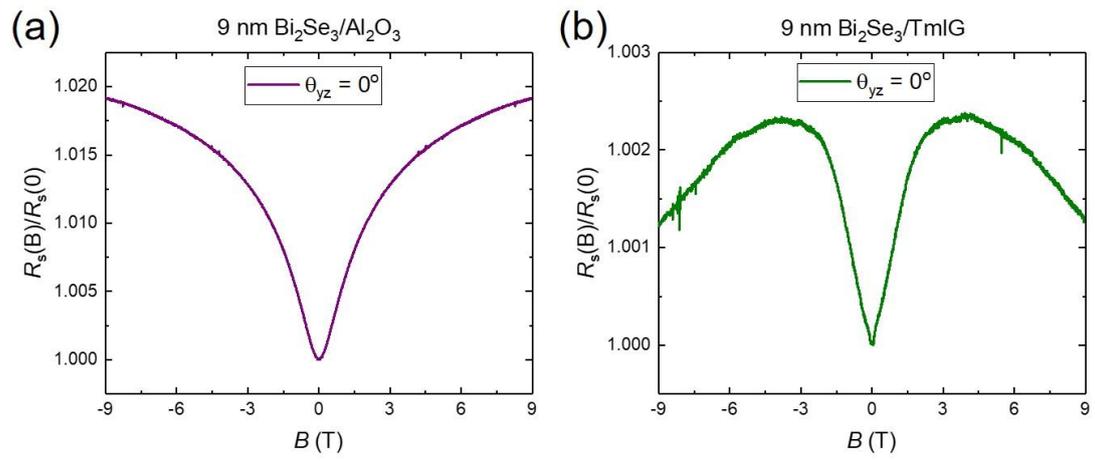

Fig. S3. In-plane MR curve of (a) 9 nm $Bi_2Se_3/Al_2O_3$ and (b) 9 nm $Bi_2Se_3$/TmIG bilayers measured at 2 K.

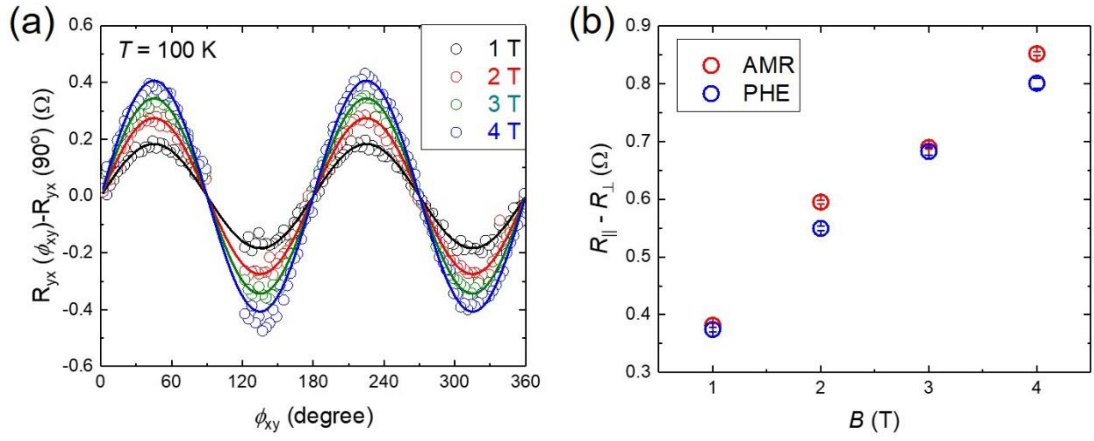

Fig. S4. Planar Hall effect data of 9 nm $Bi_2Se_3$/TmIG. (a) $\phi_{xy}$-dependent $R_{yx}$ for various $B$. Solid lines are fitted curves using Eq. (S2). (b) Comparison of the extracted AMR and PHE amplitude $R_\parallel - R_T$ as a function of $B$.

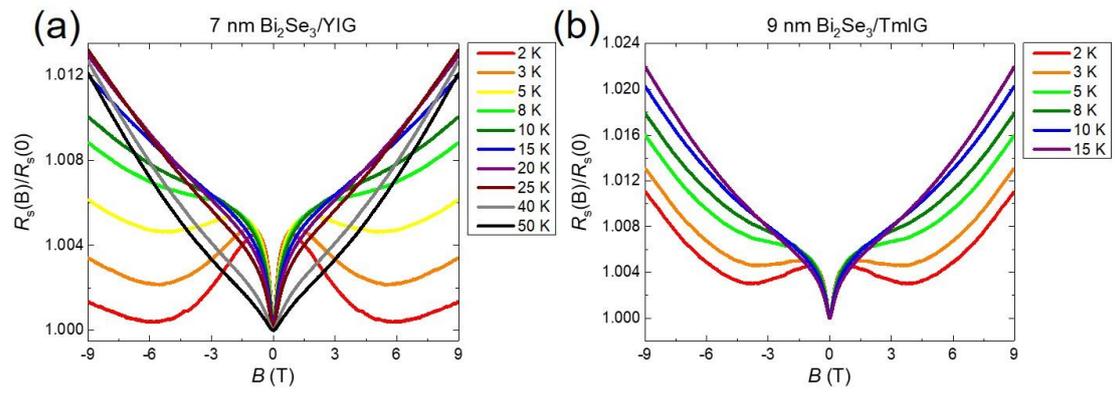

Fig. S5. *T*-dependent MR curves of (a) 7 nm $Bi_2Se_3$/YIG and (b) 9 nm $Bi_2Se_3$/TmIG.

**1. Discussions of other possible origins of the negative MR at intermediate field**

Negative MR in TI thin films could also result from physical mechanisms other than TRS-breaking of surface states. Below we describe these mechanisms and discuss their possibilities in our $Bi_2Se_3$/YIG and $Bi_2Se_3$/TmIG samples.

*(i)   Defect-induced hopping transport*

It has been shown that when crystal defects are deliberately introduced into TI thin films, a negative MR shows up as the applied field increases. As described in Ref. [37], the samples that underwent ion milling treatments show considerably modified MR characteristics. At low fields, positive MR still dominates indicating the robustness of TSSs and the WAL effect against defects. However, a negative MR starts to take over when $B > 3$ T and shows a quadratic dependence up to $B = 9$ T. The large-$B$ negative MR was attributed to field-dependent hopping probabilities among defect states in transport. In disordered semiconductors, localized spins hosted by defects lead to spin-dependent scattering of electrons. Such a process is suppressed when a large $B$ is applied that effectively aligns the localized spins, giving rise to a negative MR.

Here we note a key difference of the negative MR of $Bi_2Se_3$/YIG and $Bi_2Se_3$/TmIG from that reported in Ref. [37]. The negative MR presented in our work occurs at smaller fields (3 T $> B >$ 0.3 T), and at large fields the classical positive MR from electron cyclotron motion dominates. This is in sharp contrast to the defect-induced negative MR showing quadratic behaviors, whose magnitude is large enough to overcome the contribution of Lorentz deflection. Hence, the defect-induced hopping transport is unlikely to be the origin of negative MR in $Bi_2Se_3$/YIG and $Bi_2Se_3$/TmIG.

*(ii)   Formation of hybridization gap of top and bottom TSSs*

It is well-known that in a 3D TI, when its thickness approach 2D limits, the overlap of wave functions of the top and bottom TSSs causes a hybridization gap opened at Dirac point [48]. The 2D limit is 6 QL for $Bi_2Se_3$. The negative MR due to the hybridization gap-induced WL has been detected in bulk-insulating $(Bi_{0.57}Sb_{0.43})_2Te_3$ thin films [49]. The negative MR due to hybridization gap observed in Ref. [49] is actually similar to that in $Bi_2Se_3$/YIG and $Bi_2Se_3$/TmIG and primarily locate at even smaller $B <$ 0.3 T. The emergence of the WL effect upon hybridization gap opening can also be understood as a result of Berry phase $\pi(1 - \Delta_H/2E_F)$ deviating from $\pi$, where $\Delta_H$ denotes the size of the hybridization gap. In this aspect, both hybridization- and TRS-breaking-induced gaps give rise to WL effect. In our experiments, the WL was also detected in Bi$_2$Se$_3$ thicker than 6 QL grown on YIG and TmIG. The positive MC component can be extracted for Bi$_2$Se$_3$ as thick as 16 QL as shown in Fig. 3(c). As a comparison, the 9 QL Bi$_2$Se$_3$/Al$_2$O$_3$ sample shows a very sharp negative MC cusp of WAL. Hence, we can conclude that for Bi$_2$Se$_3$ films at 9 QL or thicker, hybridization between top and bottom surfaces should not be a concern in interpreting the WL in Bi$_2$Se$_3$/YIG and Bi$_2$Se$_3$/TmIG. Nevertheless, the similarities between the negative MR reported in Ref. [49] and our work suggests that a Dirac gap is indeed opened, despite of a completely different physical origin.

*(iii)   Quantized 2D bulk bands in thin TIs*

As studied in Ref. [38], a WL could also arise from quantized 2D subbands in ultrathin TI films. We again compare the MC of 9 QL Bi$_2$Se$_3$/Al$_2$O$_3$ and 10 QL Bi$_2$Se$_3$/YIG in Fig. 3(c). Since the two samples exhibit comparable carrier concentration $n_{2D}$, $R_s$ and thickness, the contribution of quantized bulk bands participating in transport in one sample should not differ significantly from the other. Obviously, the negative MR is absent in Bi$_2$Se$_3$ 9 QL/Al$_2$O$_3$. Hence, we rule out such subbands as the main source of the WL in Bi$_2$Se$_3$/YIG and Bi$_2$Se$_3$/TmIG.

**2.   Curve fitting details of Fig. 1**

The cusp-like feature at small fields resulted from WAL or WL is usually described by the one-component HLN equation. For our samples, the negative MR at intermediate fields implies a much larger dephasing field of WL than that of WAL. To characterize the WL component, the curve fitting range is extended, and another component of HLN equation including $\beta B^2$ (Eq. (2)) is introduced into the fitting function. It has been shown that a $\beta B^2$ term is necessary to perform the curve fitting in wide ranges of field and temperature [50]. The first two terms of Eq. (2) present the competition effects of WL and WAL. The $\beta B^2$ term account for the classical cyclotronic motion and other terms of quantum corrections in the conventional HLN equation [50]. In the following, we discuss the analyses of the MC data and the curve fitting at small and large fields separately.

*(i)   Small-field regime ($B < 1\,T$): suppressed WAL*

In this regime the weight of the $B^2$ term is negligible, so the fitting involves four independent parameters, $\alpha_0$, $\alpha_1$, $l_0$, and $l_1$ in the beginning. We found that four-parameter fittings do not render reliable results. Instead, it is possible to obtain several sets of fitted parameters that all give reasonably good fits by manually adjust the parameter values. Since WL and

WAL terms share the same mathematical form, the curve fitting is valid only when the effective phase coherence lengths $l_0$ and $l_1$ differ to some extents and the fitting range exceeds dephasing fields, otherwise an unique set of fitted parameters cannot be found. In this regime, we thus set $\alpha_0$ and $l_0$ of WL to be zero. The results are shown in the 2$^{nd}$ and 3$^{rd}$ rows of Fig. S1. Therefore, only a suppressed WAL term can be concluded in this regime for our samples.

(ii) *Larger field regime ($B > 1\,T$): negative MR and WL*

The presence of WL is seen in negative MR located at intermediate-field regime (Fig. 1(b)). To disclose the characteristics of the WL effect, the curve fitting range is extended to several Teslas. The MC curves within 4 T are shown in the first row of Fig. S1, and they exhibit a parabolic $B$ dependence toward 9 T. In this regime, five-parameter fitting, including the $\beta B^2$ term, has been performed. $\beta$ is composed of the classical cyclotronic part $\beta_c$ and the quantum correction one $\beta_q$ from the other two terms of the original HLN equation: $\beta_q B^2 \approx -\frac{e^2}{24\pi h}\left[\frac{B}{B_{SO}+B_e}\right]^2 + \frac{3e^2}{48\pi h}\left[\frac{B}{(4/3)B_{SO}+B_\phi}\right]^2$, where $B_{SO}$ and $B_\phi$ are characteristic fields of the spin-orbit scattering length $l_{SO}$ and phase-coherence length $l_\phi$. Here, $\beta < 0$ due to the negative MC at large fields.

The five-parameters fitting results are shown in the 2$^{nd}$ to 4$^{th}$ rows of Fig. S1. The MC curves of all samples can be well fitted to Eq. (2), except the one of 40 nm Bi$_2$Se$_3$/YIG which deviates the most from a 2D electron system. Since fitted parameters depend on the data range selected for the curve fitting, we display them as a function of the curve fitting range. The fitted parameters show a moderate ~10 % variations with respect to the fitting ranges. Throughout the fitting ranges, the magnitudes of $\alpha_0$, $\alpha_1$, $l_0$, and $l_1$ can be compared without ambiguity. The reliability of the five-parameter fits of the data is further justified by the following three observations. Firstly, the $\alpha_1$'s and $l_1$'s of the WAL component obtained from two-parameter fittings at small fields are in good agreement with those from five-parameter fittings. This implies the suitability of Eq. (2) for the MC behavior of Bi$_2$Se$_3$/YIG and Bi$_2$Se$_3$/TmIG in a wide range of $B$. Secondly, from the dephasing field $B_i$ calculated from $\hbar/(4el_i^2)$, we note that $B_0$ is much larger than $B_1$, which agrees with the observation that the negative MR shows up at larger fields. The notable difference of $l_0$ and $l_1$ causes the WL and WAL to manifest themselves in different regimes of $B$.

Thirdly, if we set $\beta_q \approx 0$, the electron mobility calculated by $\mu_L = \sqrt{-\beta_c R_s}$ overlap well with that calculated by our Hall measurement data $\mu_H = 1/(en_{2D} R_{xx})$, indicated by the blue dashed line. $\beta_q \approx 0$ corresponds to a very large $B_{SO}$ or small $l_{SO}$.

3. **Data of the controlled sample Bi₂Se₃/Al₂O₃/TmIG**

A trilayer sample 6 nm Bi₂Se₃/3 nm Al₂O₃/TmIG has been fabricated to test the effect of stray fields of TmIG. Here, the 3 nm Al₂O₃ layer was deposited using atomic layer deposition (ALD). The nonmagnetic Al₂O₃ insertion layer ought to suppress the interlayer exchange coupling of Bi₂Se₃ and TmIG, while allows the stray field to penetrate. Fig. S2(a) shows the MR of Bi₂Se₃/Al₂O₃/TmIG at 2 K. A cusp-like positive MR of the WAL effect was observed, and no negative MR was detected. Fig. S2(b) displays the $R_{AH}$ data as a function of $B$. No hysteresis loop was detected. Therefore, the data in Fig. S2 implies that stray fields are not the root cause of the negative MR and hysteric $R_{AH}$ loops observed in the Bi₂Se₃/TmIG.

4. **In-plane MR data at 2 K**

Fig. S3(a) and (b) show the MR data under an in-plane applied ($\theta_{yz} = 0$) field taken at 2 K for 9 nm Bi₂Se₃/Al₂O₃ and 9 nm Bi₂Se₃/TmIG, respectively. For the Bi₂Se₃/Al₂O₃ bilayer, a positive MR is detected. The MR induced by an in-plane field in Bi₂Se₃ has been studied extensively in Ref. [51]. In short, the application of in-plane fields forces the electron to scatter between top and bottom surfaces, and the presence of bulk state is essential in understanding the in-plane MR. It was demonstrated that no existing theory can well-describe the distinct transport properties of TIs under an in-plane field, thus highlighting the important role of bulk-surface coupling of TIs. The qualitatively difference between perpendicular and in-plane MR of Bi₂Se₃ can already been seen by comparing Fig. 2(d) and S3(a). While the MR with tilted field angles can be very well explained by WAL governed by $B_z$, it can be inferred that when $\boldsymbol{B}$ is rotated across a critical angle of $\theta_{yz}$ (< 30°), another physical picture of magneto-transport that dictates the in-plane MR comes into play in this regime.

For Bi₂Se₃/TmIG, the in-plane MR exhibit distinct features from those of Bi₂Se₃/Al₂O₃: the MR is positive at $B < 3$ T and becomes negative when $B$ goes larger. We may differentiate the physical origin of the in-plane negative MR from that of perpendicular one. From Eq. (1), we see that a gap in the TSS can only be opened by a perpendicular magnetization $M_z$, while $M_x$ and $M_y$ shift the gapless Dirac cone in the momentum space. Although it is argued that an in-plane magnetic field can also break TRS of TIs when the field is aligned with a certain crystal axis

[52], this should not be of importance since our $Bi_2Se_3$ films grown on garnet substrates contain randomly oriented in-plane domains. It is beyond the scope of this work to clarify the in-plane negative MR in $Bi_2Se_3$/TmIG, especially when the magnetic scattering due to MPE adds to the complexity of the system. However, we emphasize that the observation of in-plane negative MR does not pose a major problem of our interpretations of the negative MR under tilted fields. As in the case of $Bi_2Se_3$/$Al_2O_3$, a different scheme of physical model is needed for the in-plane MR.

### 5. Planar Hall effect (PHE) in $Bi_2Se_3$/TmIG

For a measurement configuration defined Fig. 2(b), the anisotropic resistivity tensor induced by an in-plane field can be reduced to two elements, $R_s$ (or $R_{xx}$) and $R_{yx}$, with respect to the sample coordinate. Phenomenologically, the field-angle-dependent $R_s$ and $R_{yx}$ are identified as AMR and planar Hall resistances, respectively when they are expressed as,

$$\text{AMR: } R_s(\phi_{xy}) = R_T + (R_\parallel - R_T)\cos^2\phi_{xy} \quad \text{(S1)}$$

$$\text{PHE: } R_{yx}(\phi_{xy}) = (R_\parallel - R_T)\sin\phi_{xy}\cos\phi_{xy} \quad \text{(S2)}$$

Fig. S4(a) shows the $R_{yx}(\phi_{xy})$ data of 9 QL $Bi_2Se_3$/TmIG. The $R_{yx}$ satisfies the angular dependence $\sin\phi_{xy}\cos\phi_{xy}$ of PHE. By fitting the data in Fig. 4(b) and Fig. S4(a) to Eq. (S1) and (S2) respectively, we extract the coefficient of the angular terms $R_\parallel - R_T$ for $R_s$ and $R_{yx}$ data. Fig. S4(b) compares the $R_\parallel - R_T$ obtained from $R_s$ and $R_{xy}$ data at various fields. One immediately sees a good agreement between the two sets of data.

PHE in TI has been previously observed in non-magnetic $(Bi,Sb)_2Te_3$ films [53] and EuS/$(Bi,Sb)_2Te_3$ [54]. In Ref. [53], the PHE results from anisotropic scattering of Dirac fermions due to TRS broken by an in-plane field. The PHE amplitude can be altered dramatically by dual-gating, showing a unique two-peak profile as the Fermi level moves across the Dirac point. We are not able to completely preclude such a scenario in $Bi_2Se_3$/TmIG bilayer, where a similar effect could also be caused by the interfacial exchange effective field. Fermi-level dependent measurements enabled by top-gating will be performed to investigate this kind of PHE. In Ref. [54], an unconventional PHE was detected, whose angular dependence cannot be described by Eq. (S2). The authors argue that a non-linear Hall response defined as $j_y = \sigma_{yxx}E_x^2$ should be considered. The proposed possible origins of the non-linear Hall response includes current-induced spin-orbit torques from TSSs, asymmetric scattering of electrons by magnons in magnetic TIs, and interband transitions between the two branches of the Dirac surface states. These scenario play minor roles, if any, in Bi$_2$Se$_3$/TmIG because such an unconventional PHE was not observed in this work.